# Compressed Sensing MRI With Variable Density Averaging (CS-VDA) Outperforms Full Sampling At Low SNR

Jasper Schoormans[1], Gustav J. Strijkers[1], Anders C. Hansen[2], Aart J. Nederveen[3], Bram F. Coolen[1]

[1] Amsterdam UMC, University of Amsterdam, Department of Biomedical Engineering & Physics, Amsterdam, the Netherlands
[2] DAMTP, Centre for Mathematical Sciences, University of Cambridge, Cambridge, United Kingdom
[3] Amsterdam UMC, University of Amsterdam, Department of Radiology and Nuclear Medicine, Amsterdam, the Netherlands

*Abstract*—We investigated whether a combination of k-space undersampling and variable density averaging enhances image quality for low-SNR MRI acquisitions. We implemented 3D Cartesian k-space prospective undersampling with a variable number of averages for each k-line. The performance of this compressed sensing with variable-density averaging (CS-VDA) method was evaluated in retrospective analysis of fully sampled phantom MRI measurements, as well as for prospectively accelerated in vivo 3D brain and knee MRI scans. Both phantom and *in vivo* results showed that acquisitions using the CS-VDA approach resulted in better image quality as compared to full sampling of k-space in the same scan time. Specifically, CS-VDA with a higher number of averages in the center of k-space resulted in the best image quality, apparent from increased anatomical detail with preserved soft-tissue contrast. This novel approach will facilitate improved image quality of inherently low SNR data, such as those with high-resolution or specific contrast-weightings with low SNR efficiency.

## INTRODUCTION

Designing MRI acquisitions always involves a compromise between scan time, image signal-to-noise ratio (SNR), tissue contrast, and spatial resolution. The development of undersampling acquisition schemes in combination with advanced reconstruction algorithms, such as parallel imaging and compressed sensing (CS) has allowed for a significant reduction in scan time, thereby minimizing image distortions (e.g. due to motion) as well as improving patient comfort and cost effectiveness of MRI protocols [1], [2]. In addition, accelerated imaging has allowed application of 3D imaging protocols at isotropic resolution, which would otherwise result in clinically unfeasible imaging times. In certain cases, however, MRI scans have inherently low SNR, e.g. when aiming for a very high spatial resolution or to achieve a specific contrast weighting, such as in T2-weighted or diffusion-weighted imaging. This limits the application of accelerated imaging as these are thought to further decrease SNR.

In recent years, the optimization of parallel imaging or CS acquisition and reconstruction techniques has received considerable attention. With respect to image reconstruction, improved transforms were designed to find the best sparse representation of the images [3]–[5], facilitating higher compression factors and thus a higher degree of undersampling. Although early work in CS theory suggests purely random subsampling of coefficients[6], much work has been performed on finding optimal sampling strategies to maximize image quality and robustness to artifacts [7]–[11], and these patterns are not uniformly distributed for MR acquisitions [12], [13].

Despite the research into the optimal distribution of sampling points, one aspect that surprisingly has received little attention is the noise sensitivity of the individual sampling points on the resulting reconstructed images. In light of the relation between wavelet and k-space coefficients [17], we hypothesized that compressed sensing reconstructions are more robust to noise disturbances in high frequency regions. In this work, we therefore aimed to show that employing an undersampling and averaging scheme (without affecting total acquisition time) results in superior image quality as compared to full sampling without averaging.

While this seems unintuitive at first, this study clearly shows the advantage of this approach in a number of steps. First, the rationale and implementation of our new acquisition method of Compressed Sensing with variable averaging (CS-VDA) is presented, combined with a noise-optimal weighted l2-norm in the CS reconstruction. It can be shown that the weighted l2-norm, considering the number of averages for every line, is a least-squares optimizer for this reconstruction problem. In simulations, we were able to show that our new CS-VDA approach, given equal total scan time, provides superior image quality to fully sampled data. To further demonstrate the performance of our CS-VDA approach, we performed experiments using prospectively undersampled *in vivo* brain MRI scans, as well as quantitative T2 MRI of the knee, using clinical 3T MRI.

## THEORETICAL BACKGROUND

In this paper, we introduce a new compressed sensing sampling method with variable averaging (CS-VDA). Assume we have a fixed sampling pattern, consisting of $n$ k-points, but a scan-time budget of $m>n$ sampling points, *i.e.* we must distribute these $m$ points over $n$, by resampling points in some way. Figure 1 shows three ways to distribute sampling points: by uniform averaging, center-dense averaging; and periphery dense-averaging. For all experiments in this work, k-space was undersampled.

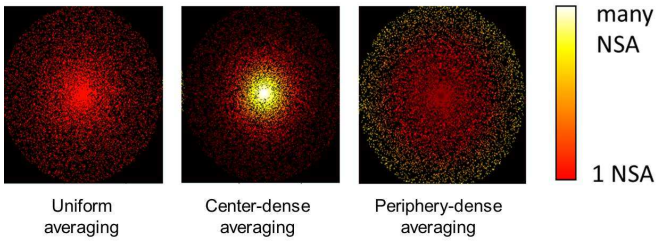

*Figure 1. Illustration of the three k-space undersampling & averaging strategies. Shown are the 2 phase encoding dimensions of the 3D k-space; the 3rd dimension is the frequency encoding direction. The k-spaces share the same variable density undersampling pattern, with denser sampling in the center of k-space than in the periphery. However, the three k-spaces differ in how the number of signal averages (NSA) are distributed in k-space, with (A) uniform averaging of all measured k-space points, (B) higher NSA in the center of k-space (center-dense averaging), and (C) periphery-dense averaging. Note that all strategies have the same total number of k-space points.*

To obtain a better understanding of the effect of the proposed sampling strategy, we first illustrate the influence of noise in the wavelet domain, a commonly used sparsity transform in MRI compressed sensing. Figure 2 shows the wavelet transform of the Shepp-Logan phantom, with complex noise of an equal $l_2$-norm added in the two following ways: A) the noise was added in the 5 lowest-frequency levels of the wavelet coefficients; B) the noise was added to the highest level of wavelet coefficients only. From the resulting inverse wavelet transforms, it is evident that the addition of noise to the low wavelet levels has a much worse effect on the general quality of the image: the details indicated by the red arrow are much more difficult to distinguish in this case.

While the wavelet transform is commonly used to employ sparsity in compressed sensing, MR images are acquired in k-space. Fig. 3 shows the Shepp–Logan phantom in these two transforms. The MR measurement of a wavelet signal in Cartesian k-space is the subsequent operation of the inverse wavelet transform (DWT*) and the discrete Fourier transform (DFT), and can be formulated as the matrix $U= DFT\ DWT^*$.

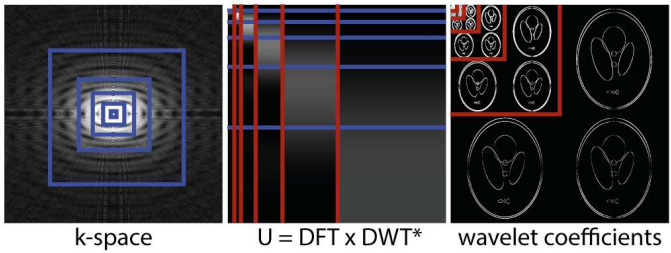

k-space     U = DFT x DWT*     wavelet coefficients

*Figure 3 The structure of the MR measurement. The MR measurement of this wavelet signal in Cartesian k-space is the subsequent operation of the inverse wavelet transform (DWT*) and the discrete Fourier transform (DFT), and can be formulated as the matrix U= DFT DWT*. The absolute values are displayed here. Wavelets of increasing coefficients are indicated by red lines, and increased frequencies in k-space are illustrated by blue lines.*

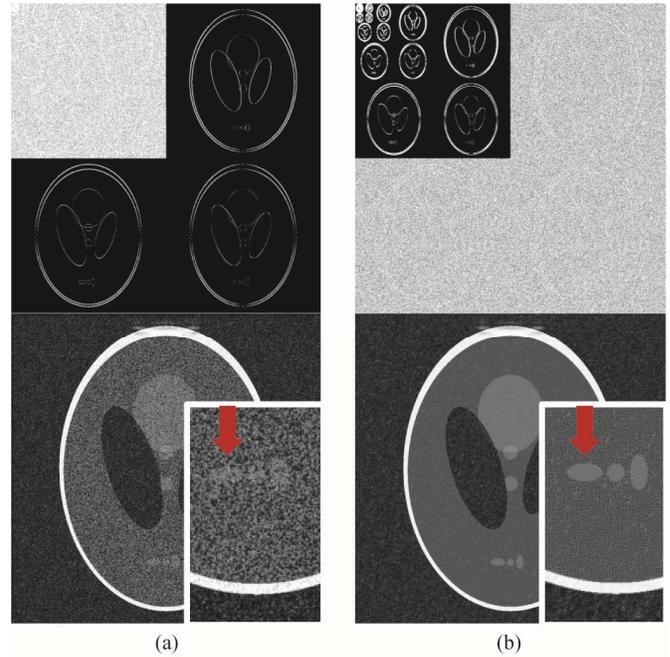

*Figure 2. Influence of noise in the wavelet domain on image reconstructions. A) Noise was added to the coarse wavelet coefficients. In B), noise with an equal $l_2$-norm as in A. is added to the highest wavelet level only. It is clearly visible that resulting details in the reconstruction (red arrow) are much more easily distinguished in B.*

This matrix is nearly block-diagonal, with each row representing a k-point measurement. The wavelet coefficients exist in increasing scales, each scale illustrated by a red bounding box. The sections of matrix $U$ acting upon each separate wavelet scales are distinguished by the red lines. In contrast to wavelets, the frequencies in k-space are linearly increasing – not fixed blocks. However, if we show increasing resolutions by blue boxes (1mm, 2mm etc.), this converts to the blue lines in $U$. Now, because measurement $U$ is nearly block-diagonal, we can say that wavelets at a given scale are essentially concentrated in square rings of k-space.

Figure 4 shows a one-dimensional toy example, combining the insights from Figs. 2 and 3. Consider a one-dimensional object to be measured: the projection of the Shepp-Logan phantom. Resembling many natural signals, this signal has a sparse representation in the wavelet domain (Fig. 4B). Moreover, it shares the sparsity structure of most natural images: the wavelet coefficients are unequally distributed, with most of the signal energy made up from coarse wavelets, and few non-zero coefficients lying in the higher coefficients. Consider two resampling strategies: we will measure all k-points at least once, but we can choose to measure either the low-resolution (Fig. 4C; case 1), or the high-resolution half (Fig. 4D; case 2) of k-space many times, such that there is effectively no more noise in that half. In the wavelet domain this will result in the following: case 1) noise mostly concentrated in the higher wavelet coefficients encoding for the details; case 2) noise mainly in the lower wavelet coefficients, which encode for coarse image structures and contrast.

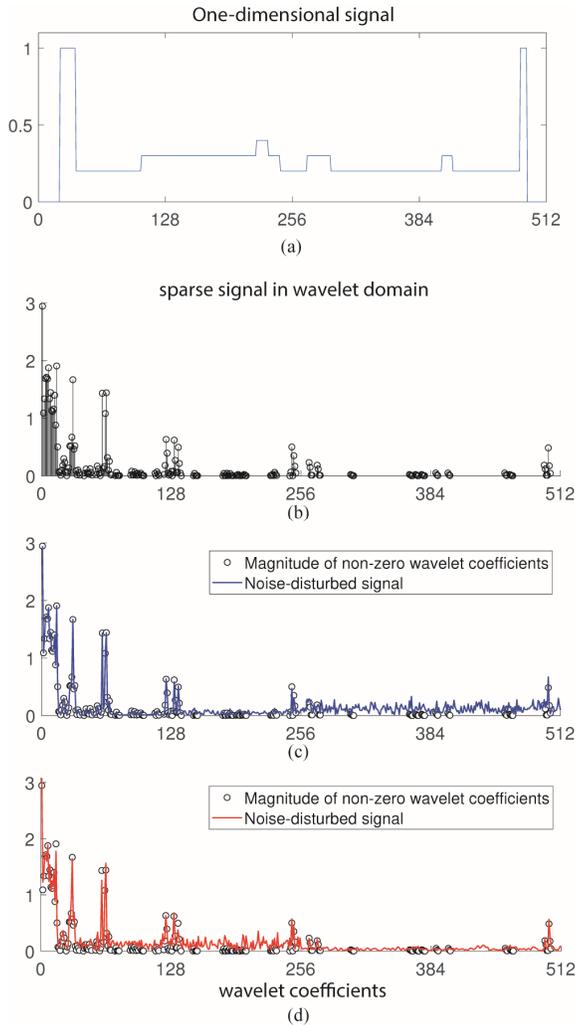

**Figure 4**. *The structure of noise in a 1D MR measurement. A) A one-dimensional signal. B) The wavelet coefficients of this signal. C) Noise disturbed wavelet coefficients for infinite resampling in the low-frequency half of k-space. D) Noise disturbed wavelet coefficients for infinite resampling in the high-frequency half of k-space.*

This has the following effects: firstly, as was illustrated in Figure 2, the image quality is considerably worse in the case of coarse-wavelet noise. Secondly, because $U$ is nearly block-diagonal, when the noise is concentrated on the lower wavelet coefficients (case 2), it is projected onto few coefficients in k-space. In contrast, for case 1 the noise contribution will be shared with many more points. Along the same lines as in Adcock *et al.*[18], where the structure of $U$ is the basis of the explanation of the variable-density sampling strategy in CS, this structure can also point in the direction of success in designing an averaging strategy. Thirdly: while one of the key tenets of CS is incoherence, it has been shown that the MRI measurement matrix is only asymptotically incoherent: In the low-resolution regime, the measurement is very coherent, as indicated by the concentrated coefficients in the columns of $U$. For higher resolutions, $U$ becomes increasingly more incoherent. Given these considerations, we theorize that higher frequency k-points are more robust to noise, as the CS theory is more applicable in this regime.

It is evident that the toy example given in Fig. 4 is extreme: the examples given correspond to the unrealistic limit of infinite resampling. A full theoretical explanation of the effects, given realistic sampling numbers, is outside the scope of this paper. However, the behavior described in these toy examples suggests that resampling of points should be focused on the low resolution, coherent, k-space. **Therefore, we predict that for low SNR measurements, unevenly distributing averages with more averages in the low frequencies of k-space will result in superior image quality, even compared to full sampling.** In this paper, we have investigated three types of averaging distributions (Figure 1). A full investigation over the range of distributions, and its dependence, on SNR, resolution, sparsity structure etc. is outside the scope of this work. This paper introduces the concept, and argues for CS-VDA in certain low-SNR applications. The remainder of this will work focus on simulations and prospective measurements of the CS-VDA strategy.

**Acquisition strategy**

For all experiments in this work, a randomly generated variable density [1] undersampling pattern was generated, according to the probability density function

$$P(r_k) = c + (1 - r_k)^4, \quad (1)$$

where $r_k$ is the distance to the k-space center. The scaling factor $c$ was chosen such that

$$\int P(r_k) dr_k = \frac{N_k}{R}, \quad (2)$$

*i.e.* the total number of k-space samples for a fully sampled scan $N_k$ divided by the undersampling factor $R$. In every undersampled scan, the total number of measured k-space samples was kept identical: $N_k$. This implies that the total acquisition times of undersampled and fully sampled scans were equal. Thus, for the undersampled scans, k-space points could be averaged since there are more readouts than available k-space positions. This averaging in k-space was done in one of the following three ways (Fig. 1):

i) *Uniform averaging*
   Every sampled k-space point was averaged an equal number of $R$ times, with $R$ an integer number.
ii) *Center-dense averaging*
   Averaging was denser in the k-space center, *i.e.* more averages were taken in the center of k-space, and fewer in the periphery. The number of signal averages (NSA) was determined using the sampling probability density

$$\text{NSA}(r_k) = N_{\max}(c + (1 - r_k)^p), \quad (3)$$

where $p$ determines the distance-dependent sampling. In this work, $p$ was chosen to be 4, $c$ similar as the sampling distribution as in Equation 1, and the scaling factor $N_{max}$ such that the total number of acquisitions equals

$$N_k = \int \text{NSA}(r_k) P(r_k) dr_k. \tag{4}$$

iii) *Periphery-dense averaging*

Fewer averages were taken in the center, and more in the periphery, according to

$$\text{NSA}(r_k) = \frac{1}{\beta + (1 - r_k)^p}, \tag{5}$$

where $\beta$ is a normalization constant calculated with Eq. 4. For both center-dense and periphery-dense averaging NSA($r_k$) was rounded off to its nearest integer number.

**The weighted l$_2$-norm**

Noise in k-space can be modeled as a normally distributed stochastic variable $N(0, \sigma^2)$ with zero mean, and a standard deviation $\sigma$ independent of k-space location [19]. The maximum likelihood estimator of independent and identically normally distributed data is the least-squares estimator, which is related to the data fidelity term

$$|F_u S m - y|_2^2 = (F_u S m - y)^*(F_u S m - y), \tag{6}$$

where $F_u$ is the undersampled digital Fourier transform operator, is a matrix containing the coil sensitivities, $m$ is the image vector, and $y$ is the multichannel undersampled k-space. However, in the case of non-uniform averaging in k-space, the assumption of identical noise variance per k-space point is violated. In this case, the maximum likelihood estimator is given by the weighted least-squares estimator

$$|F_u S m - y|_{2,W}^2 = (F_u S m - y)^* W (F_u S m - y), \tag{7}$$

where $W$ denotes the variance-covariance matrix. Assuming independent noise, the covariances are zero. From the sample mean, we can estimate the variances

$$\sigma_i^2 = \frac{\sigma_0^2}{n_i}, \tag{8}$$

where $n_i$ is the number of averages for the $i^{th}$ k-point, and $\sigma_0$ is the variance for a single average. The weighting matrix then becomes

$$W = \frac{1}{\sigma_0^2} \begin{bmatrix} n_1 & \cdots & 0 \\ \vdots & \ddots & \vdots \\ 0 & \cdots & n_N \end{bmatrix}, \tag{9}$$

where $N$ is the total number of measured k-points. Of note, a similar weighting matrix in the l$_2$-norm has been proposed previously for general noise uncertainties [20]. In this work, we use the weighted l$_2$-norm in combination with the l$_1$-norm, according to

$$\widehat{m} = \arg\min_m \left| W^{\frac{1}{2}}(F_u S m - y) \right|_2^2 + \lambda |Q m|_1, \tag{10}$$

with $\lambda$ a regularization parameter and $Q$ a sparsifying transform, *e.g.* the wavelet transform or a finite-difference operator.

**Noise variance and the l$_2$-norm**

The sampling patterns described in Equations 3,4,5 will influence the noise variance and lead to different data fidelity terms. To keep the balance with the l$_1$-norm. the λ in Equation 10 is adapted to the expected l$_2$-norm:

$$\lambda_{mod} = \alpha \, \lambda_0 \tag{11}$$

where α is the regularization adaption term.
The expected data fidelity in Equation 10 is given as:

$$\mathbb{E}(l_2^2) = \mathbb{E} \sum_{i=1}^{N} \left| W_i^{\frac{1}{2}} (y_i - y_{i,0}) \right|^2, \tag{12}$$

Where $y_i$ is the averaged measurement of the $i^{th}$ k-line, and $y_{i,0}$ is the true value of that k-line. This simplifies to:

$$\mathbb{E}(l_2^2) = \sum_{i=1}^{N} \mathbb{E}\big(W_i (y_i - y_{i,0})^*(y_i - y_{i,0})\big), \tag{12}$$

and simplifies further to:

$$\mathbb{E}(l_2^2) = \sum_{i=1}^{N} W_i \cdot \Big( \mathbb{E}(y_i^* y_i) + y_{i,0}^* y_{i,0} - 2 y_i^* \mathbb{E}(y_i) \Big). \tag{13}$$

As we assume white noise, the expected values of the measurement $\mathbb{E}(y_i) = y_{i,0}$ and $\mathbb{E}(y_i^* y_i) = y_{i,0}^* y_{i,0} + |\sigma_i|^2$, Equation 12 becomes

$$\mathbb{E}(l_2^2) = \sum_{i=1}^{N} W_i \cdot |\sigma_i|^2 = N. \tag{14}$$

From the definition of the weighting matrix in Equation 9, the regularization is then be scaled with:

$$\alpha = \frac{\mathbb{E}(l_2^2)}{\mathbb{E}(l_2^2)_{full}} = \frac{N}{N_0} = \frac{1}{R}, \tag{15}$$

where $N_0$ is the number of samples for a fully sampled, once-averaged measurement.

# METHODS

**Retrospective undersampling of phantom data**

MRI was performed on a 3T clinical MRI scanner (Philips Ingenia, Best, the Netherlands). For retrospectively undersampled experiments, we chose to scan a grapefruit phantom to maximize the presence of detailed structures. A high-resolution 2D $T_1$-weighted fast-field echo scan of the grapefruit was performed using the following parameters: flip angle (FA) = 15°, TR = 15.6 ms, TE = 4.2 ms, FOV= 128×128 mm$^2$, matrix size = 512×512, resolution = 0.25×0.25 mm$^2$, and slice thickness = 1 mm. This fully sampled scan was made with 50 averages for every k-space point to allow full flexibility in retrospective undersampling of k-space points and NSA per point. This dataset was retrospectively undersampled and averaged for $R$ = 1, 2, 3, 4, and 5 in the above described three ways. Four different regularization parameters $\lambda$ were tested (0.0005, 0.005, 0.05, and 0.5). After reconstruction, images were normalized with the *normalize* function in the BART toolbox [21]. Scan quality was assessed by fitting a sigmoid perpendicular to the air/fruit-skin interface, with the Matlab *lsqnonlin* fit routine. For every reconstruction, this was done for ten lines. The sigmoid width parameter was used as a measure for apparent sharpness [22].

To further investigate the effect of the weighting matrix, a second scan of the grapefruit was made with the same settings as described above. The k-space data was subsampled 5 times with a center-dense variable density pattern, with variable averaging ranging from 25 averages in the center to 5 in the periphery of k-space. This data was reconstructed in 4 different ways, *i.e.* using an l$_2$-norm, a weighted l$_2$-norm, an l$_2$-norm plus the l$_1$-norm, and finally, our proposed weighted l$_2$-norm combined with an l$_1$-norm (Eq. 10). For both methods, 30 iterations of the non-linear conjugate gradient algorithm were used and the optimal $\lambda$ was determined by the l-curve method [23]. The noise spectral density was calculated in a noise-only region-of-interest, outside the fruit.

**Prospective undersampling of in vivo human brain data**

The institutional review board of our hospital approved this study. All four healthy volunteers (2m, 2f, age=26-35) gave written informed consent for participating in this study. For prospective undersampling of k-space, an in-house developed scanner software patch was used to sample user-defined k-space trajectories. An inversion prepared 3D $T_1$-weighted fast-field-echo brain scan was performed at high isotropic spatial resolution using a 16-channel head coil and the following sequence parameters: FA = 5°, TR = 7.9 ms, TE = 2.6 ms, inversion delay time (TI) = 1000 ms, echo train length = 120, FOV= 210×210×58 mm$^3$, matrix size = 304×302×91, resolution = 0.7×0.7×0.7 mm$^3$. For each volunteer, five different k-space patterns were measured, *i.e.* one fully sampled k-space, uniform averaging with either $R$ = 3 or 5, and center-dense averaging with $R$ = 3 and 5. Total scan time was 6min24s for all scans. Furthermore, a fully sampled lower resolution acquisition with a resolution of 1×1×1 mm$^3$ and uniform NSA = 2 was acquired. Scans were assessed visually for image sharpness and signal-to-noise.

**Prospective undersampling of in vivo human knee data**

A T2-prepared fast-field echo knee scan of a healthy female volunteer (26y) was performed at 3T, using a 16-channel knee coil. The sequence was adapted from Colotti *et al.* [24], modified with selective water excitation, and a segment-time increased from 700 to 800 ms. Fully sampled and CS-VDA ($R$ = 3) data were scanned in an interleaved fashion, for T2-

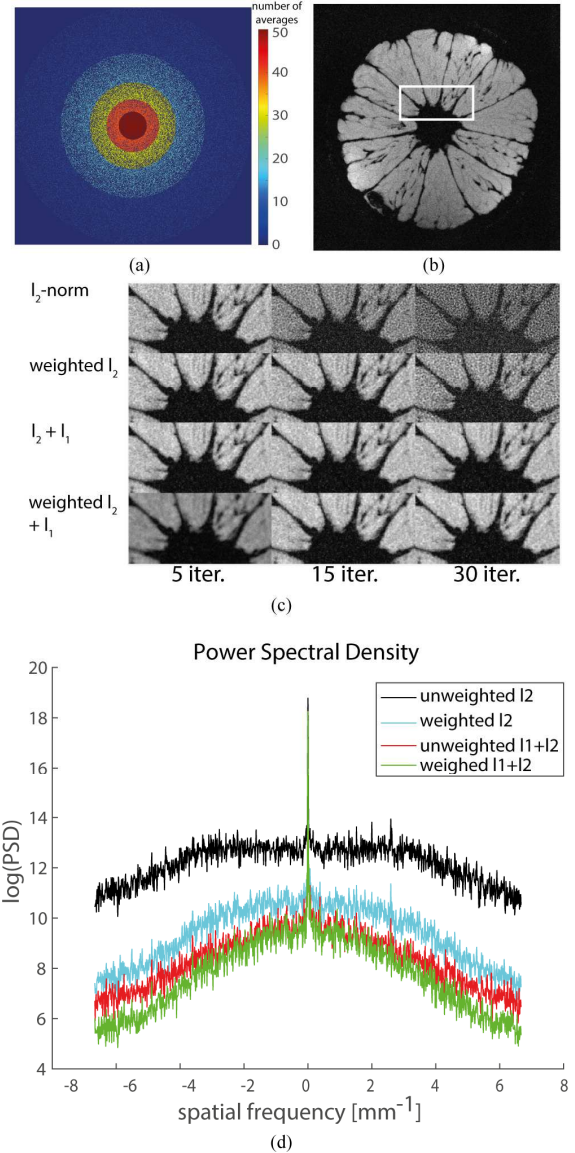

*Figure 5* Imaging of grapefruit using different reconstruction methods. (A) Center-dense undersampling of k-space. (B) Middle slice of a fully sampled scan with NSA = 50. (C) Magnifications of the regions of interest, shown in A, with different reconstruction strategies and number of iterations. (D) Noise power spectral density for the four methods. The use of the noise weighting matrix reduces the power spectral density over the full range of spatial frequencies, both without inclusion of l$_1$-norm regularization (black to blue) and with l$_1$-norm regularization (red to green).

echo preparation times of 0, 23, 38 and 58 ms. Further scan parameters were: FOV = 140x150x171 mm$^3$, resolution: 0.8x0.8x0.8 mm$^3$, FA = 15 degrees, TE/TR = 3.4/6.9 ms, TFE-factor = 100. Total scan time amounted to 4 x 4m51s per sampling method. The reconstruction parameter $\lambda$ was optimized for every separate T2-prepared acquisition: a value of λ=0.001 was used for TE = 0, and λ = 0.005 for all other scans. Three-dimensional rigid registration and a pixel-wise T2 fits were performed with Matlab (R2015b).

**Image reconstruction**

All image reconstructions were performed in Matlab (R2015b, The MathWorks, Natick, 2014). Both full and undersampled data went through the same reconstruction pipe-line. K-space data was loaded and preprocessed with the MRecon toolbox (version 5.3.19, Gyrotools, Zürich, Switzerland). The data was pre-whitened and sensitivity maps were calculated with the ESPiRIT method in the BART toolbox [25][21]. Slices were reconstructed in parallel after an iFFT in the frequency encoding direction. Equation 10 was solved by a non-linear conjugate gradient algorithm. The number of iterations was carefully chosen by visual comparison to prevent noise amplification [26] and a restart strategy with 3 outer iterations was used. A wavelet transform was used as a sparsifying operator for the *in vivo* scans, and a total-variation constraint was used for the grapefruit scans. Reconstruction time was 5 minutes for a full reconstruction of a 3D dataset on a standard Dell workstation (3.5 GHz, 32 GB memory), with a Geforce Titan XP GPU.

# RESULTS

Figure 5 illustrates the beneficial effect on image quality of the weighted l$_2$-norm in comparison to the traditional unweighted l$_2$-norm. Figure 5B shows the middle slice of the fully sampled 3D dataset of the grapefruit with NSA = 50 (experiment 1), providing a reference to which the reconstructions with the undersampled and center-dense NSA k-space pattern (Fig. 5A) can be compared. Magnifications of the regions of interest in Fig. 5B with different reconstruction strategies and iteration numbers are shown in Fig. 5C. Upon visual inspection, the reconstructions with l$_2$-norm + l$_1$-norm and weighted l$_2$-norm + l$_1$-norm (Eq. 10) with 15 and 30 iterations resulted in the best image quality with the lowest noise. This visual assessment can be objectified by comparing the power spectral density calculated from a background-only (noise) region in the image for the different reconstruction methods (Fig. 5D). The use of the noise weighting matrix $W$ reduced the noise power over the full range of spatial frequencies. In CS reconstructions that included the l$_1$-norm, the benefit was visually less apparent, but still substantial as is shown in Fig. 5D.

Figure 6A shows half of a center slice of the grapefruit, reconstructed from five times undersampled ($R$ = 5) data using uniform, center-dense, and periphery-dense averaging in comparison to the fully sampled scan. For a fair comparison, all reconstructions, including the fully sampled scan, were performed using Eq. 10 with $\lambda$ = 5×10$^{-4}$ and equal total number of samples. In Fig. 6B, reconstructions are shown for varying regularization parameter $\lambda$. The width of the sigmoid-curves fitted to the air/grapefruit-skin interface (Fig. 6C) is lowest, *i.e.* the interface is sharpest for the center-dense sampling scheme with $\lambda$ = 5×10$^{-3}$ and $R$ = 4. This approach results in a sharper interface than the corresponding fully-sampled scan, approaching the sharpness of a very high-SNR reference. Only for the higher values of $\lambda$, fully sampled scans are sharper than the undersampled counterparts, but this is due to over-smoothing of all images (Fig. 6B).

For the *in vivo* experiments, both a CS-VDA brain scan with center-dense averaging ($R$ = 3), as well as a fully sampled scan were acquired at 0.7 mm isotropic resolution. In addition, a low-resolution (1 mm isotropic) fully sampled scan was acquired. All acquisitions were performed in equal total scan time. Because of the – deliberately chosen – low flip angle of the read-out train, images resulting from the fully sampled scan were noisy with almost no visible anatomical details (middle row, Fig. 7). In comparison, the CS-VDA scans with center-dense averaging showed considerably better signal-to-noise (top row, Fig. 7) and clearly displayed detailed anatomical features such as vessels and outlines of sulci. To prove that the improved signal-to-noise and image sharpness was due to the specific k-space sampling pattern with center-dense averaging and not merely a signal smoothing effect, we also scanned with lower resolution in the same scan time (bottom row, Fig. 7). As expected, lower resolution scans did not present the level of anatomical detail as was seen for the images acquired with the CS-VDA approach.

Figure 8A shows scans of one volunteer for varying degrees of undersampling (full sampling, $R$ = 3 and 5) and magnifications in three orientations. Corresponding k-space sampling patterns are shown in Fig. 8B. Again, fully sampled scans at low and high resolution as well as scans with uniform k-space averagingwere noisy and lack anatomical detail. However, the variable density averaging images have much better image sharpness and signal-to-noise. For $R$ = 5 though, images appear slightly smoothed, particularly for the coronal cross section. Figure 9 shows results from the T2-mapping knee acquisitions. The images in Fig. 9A corresponds to the highest T2-prepared echo time (TE = 58 ms), which has the lowest SNR. Red arrows indicate detailed structures of cartilage and muscle that are recovered in CS-VDA, while appearing not sharp (or lost in the noise) in the fully sampled acquisition. Fig. 9B shows the loss of details in a zoomed-in section, corresponding to the red box in subplot A, occurring at the later echo times in the fully sampled acquisition. This leads to underestimation of T2, as shown in Fig 9C. Finally, the calculated SNR values of a ROI in the muscle reveals an increased SNR for CS-VDA, for all acquired echo times, as well as for the reconstructed T2-map. The calculated T2 for muscle was (mean ± standard deviation) T2 = 26.1±0.9 ms (CS-VDA); T2 = 25.0±1.6 ms (fully sampled). Supplementary figure S1 provides an animated gif, covering a range of slices of the same acquisition.

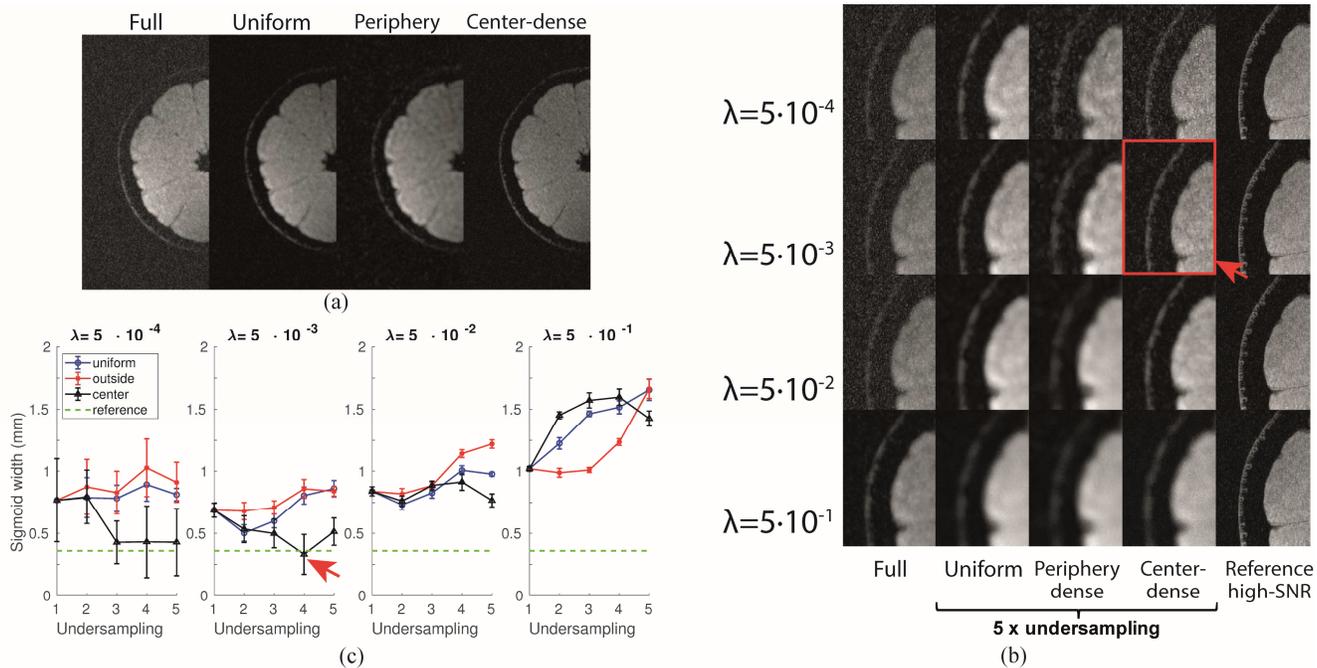

*Figure 6.* Comparisons of different CS-VDA sampling schemes. (A) half of a center slice of the grapefruit, five times undersampled (R = 5), reconstructed using uniform, center-dense, and periphery-dense k-space averaging in comparison to the fully sampled scan. (B) Enlarged sections of the reconstructed images with varying regularization parameter λ. Note the thin layer covering the grapefruit skin, which is sharpest for the center-dense averaging scheme with λ = 5×10$^{-3}$. Undersampled scans have better signal-to-noise than the fully sampled scan. (C) Mean calculated sigmoid widths for different sampling and reconstruction parameters. The fully-sampled scan corresponds to undersampling = 1. The error-bars indicate the standard deviation of 10 measurements. The two lowest sigmoid widths, and the corresponding images are indicated by red arrows.

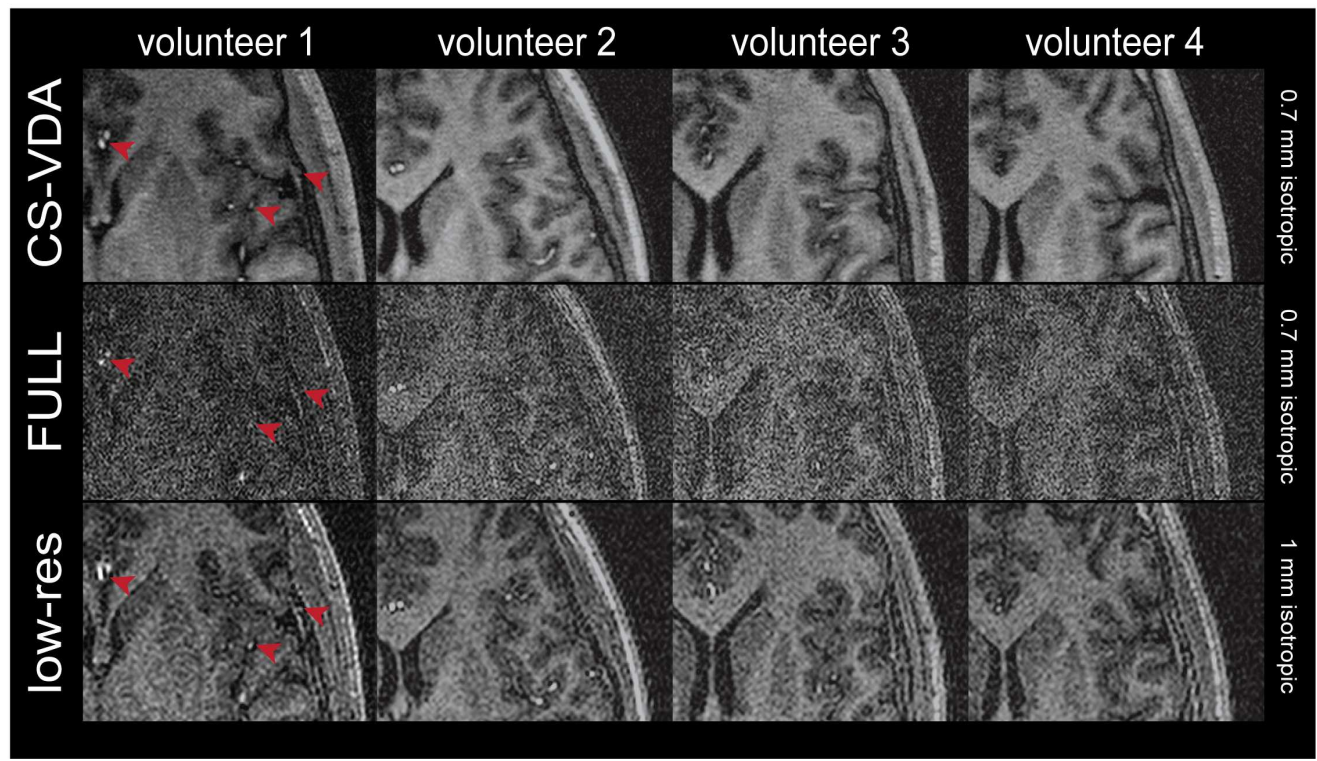

*Figure 7.* Inversion prepared 3D T1-FFE brain scans of the four volunteers scanned with (top) center-dense averaging (R = 3) and 0.7 mm isotropic voxel size, (middle) fully sampling at 0.7 mm isotropic voxel size, and (bottom) full sampling at 1 mm isotropic voxel size. All scan times were equal. The red arrows indicate small anatomical details in the brain.

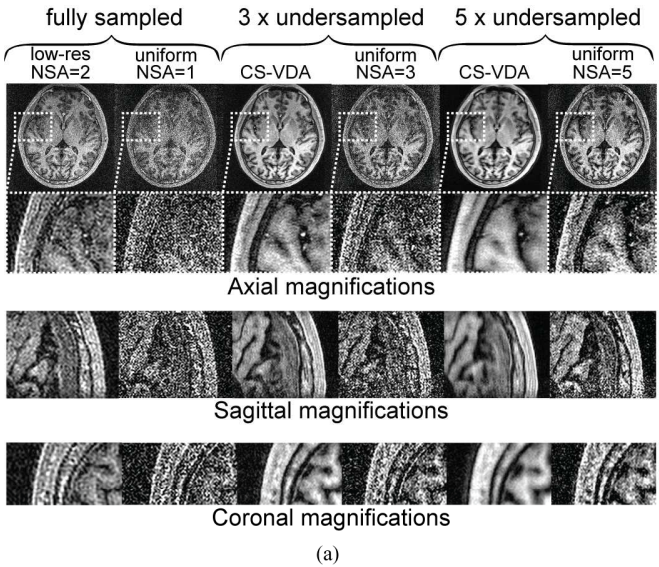

Axial magnifications

Sagittal magnifications

Coronal magnifications

(a)

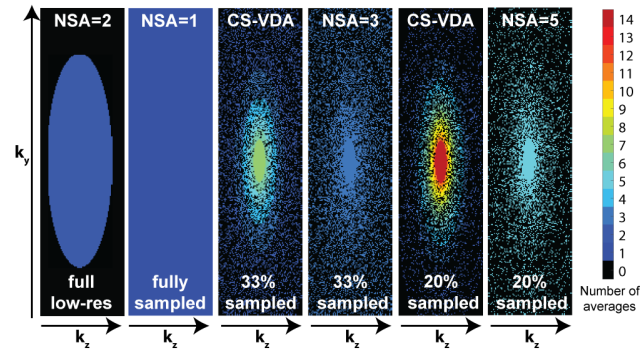

(b)

*Figure 8. Inversion prepared 3D T1-FFE brain scans of a single volunteer acquired with 6 sampling schemes of equal total scan times. (A) Reconstructed images with magnifications in the three orientations. (B) Corresponding k-space sampling patterns. Colors indicate the number of averages for every k-line.*

## DISCUSSION

In this study, we demonstrated that image quality in SNR-deprived volumetric scans can be improved by using both k-space undersampling and averaging, combined with CS reconstruction while maintaining the same total scan time as a fully sampled scan. We introduced and tested three different undersampling and averaging methods. The averaging strategies were: uniform averaging; more averages in the center and more averages in the periphery of k-space. We found that in terms of image quality, most benefit was gained by center-dense averaging. In Fig. 6, the three different strategies are employed for a retrospectively undersampled acquisition of a grapefruit. The measured thickness of the fruit skin, a surrogate for image sharpness, was lowest for the center-dense averaging at four times scan acceleration. Uniform averaging also impacted the resolution positively, compared to full sampling with only one average.

Deterministic variable averaging has been used as an alternative to a low-pass filter [14], [16], with an SNR benefit that was found to be 17 percent in 31P-MRI [15]. The deterministic approaches do not average multiply sampled points, but rather sum the points as to achieve low-pass filtering to reduce Gibbs ringing. In contrast to these earlier works, we do average all sampled points, and combine it with a compressed sensing acquisition. Instead of Gibbs ringing removal, we achieve image quality improvements explained by the noise characteristics and the wavelet-Fourier relation. The link between wavelet-domain sparsity and k-space was used to design the Subband Compressed Sensing with Quadruplet Sampling method [17], where parallel imaging was used in the low-frequency k-space. As opposed to CS-VDA, this method requires high SNR and high contrast.

Since variable k-space averaging introduces a non-uniform noise variance, we included a weighted $l_2$-norm in the image reconstruction minimization function. This led to a significant improvement in image quality, in agreement with Johnson *et al* [20] who introduced the $l_2$-norm weighting to improve image reconstructions of data with unfavorable contrast evolutions in k-space. In the presence of an $l_1$-norm, the noise reduction gain when using a weighted $l_2$-norm was smaller but still significant and therefore we recommend the addition of the weighted $l_2$-norm for these types of acquisitions. Note that for uniform sampling, the added weights have no effect, since they are the same for all k-space lines. At low iteration numbers, the weighted $l_2$-norm + $l_1$-norm for the center-dense acquisitions (Fig. 5C) resulted in more blurring as compared to the non-weighted versions. The reason for this is that at the start of iterative reconstruction – when convergence has not been reached yet – there is more weight on the center of k-space. This initially leads to some blurring, which is resolved at higher iteration numbers.

In our experiments, the number of signal averages was based on a power function (Eq. 3). The optimal distribution of averages throughout k-space will depend on several factors, such as the specific sampling distribution, noise level, matrix size, etc. Such optimization may be a topic of future research. Supplemental Figure 2 shows the effect of varying p on different image quality measures for the R = 4 case in Figure 6. These results justify the choice for *p* = 4 in Eq. 3 to create the center-dense averaging distribution.

As an *in vivo* proof of concept, we applied the center-dense averaging strategy to high-resolution 3D imaging of the human brain (Figs. 7 and 8). The quality of images acquired with center-dense averaging method was significantly better than fully sampled images acquired in the same scan time. Image quality in terms of sharpness and signal-to-noise was also better than fully sampled images at lower resolution, which hows that center-dense averaging is not equivalent to low-resolution high SNR imaging. Nevertheless, the distribution of points should remain balanced between the k-space center and periphery and we have observed that *R* = 4 resulted in the sharpest images (Fig. 6). While the proposed approach is well-suited to counterbalance decreased SNR of high-spatial resolution acquisitions, effective resolution can be decreased as a result of motion. To achieve true high resolution images, strategies to minimize physical motion or apply prospective motion correction could be considered [27]. Although we did not investigate this explicitly, an additional benefit of center-dense averaging could be increased motion robustness. While this might explain some of the quality improvement in the brain images, the sharpness improvements in retrospective grapefruit scans, which were not influenced by any motion, show that this is not the only factor.

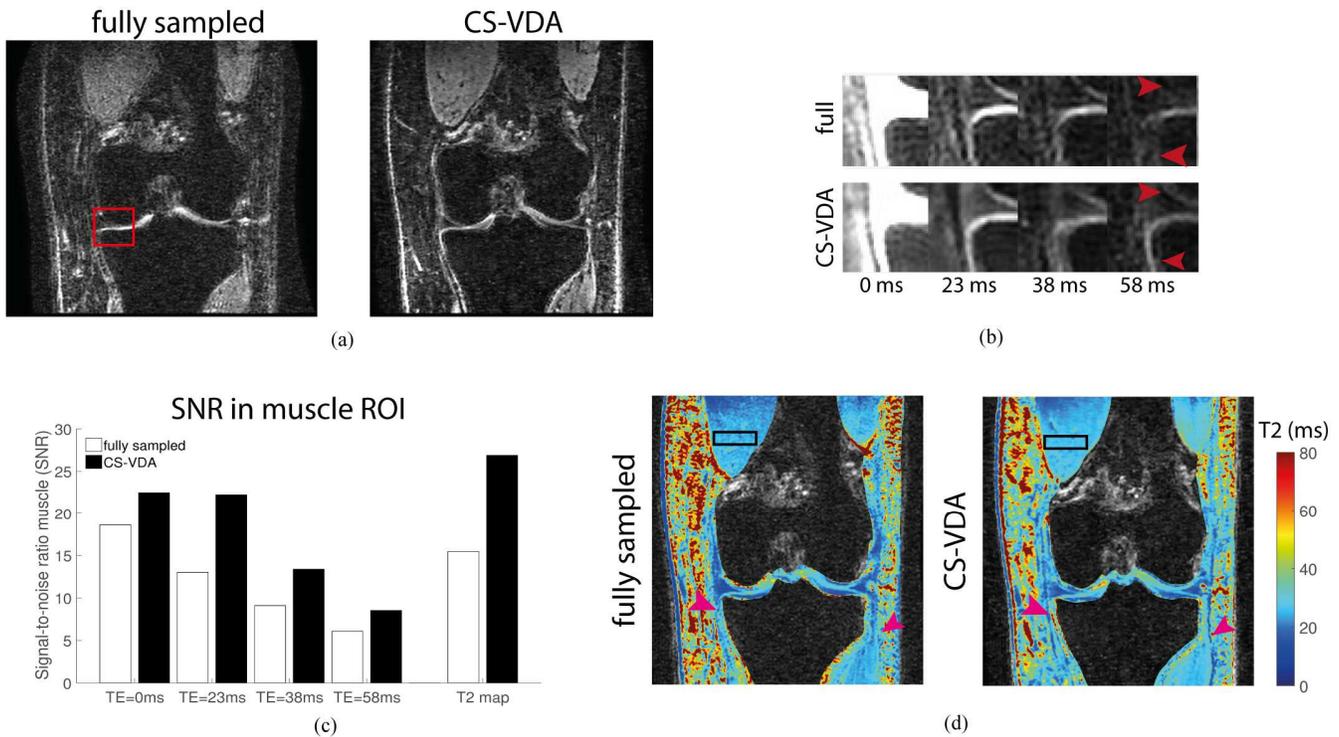

*Figure 9. T2 mapping experiment. (A) Coronal slice of a T2-prepared acquisition with TE=58 ms. Red arrows indicate loss of details in the fully sampled acquisition, with respect to the CS-VDA acquisition (B). Zoomed-in section of image for four different echo times, corresponding to the red arrow with the white star. Loss of structural detail in the fully sampled acquisitions is apparent for TE=38 ms and TE=58 ms. (C) Signal-to-noise ratio in a section of the muscle (box of subplot a). For all echo times and for the T2 map, the SNR is higher in CS-VDA compared to full sampling (D) T2-maps for the cartilage. Arrows indicate regions of difference between the two acquisitions: the high T2 of fluid and the delineation of the muscle is represented more clearly in the CS-VDA T2 map.*

We here again want to explain that 3D brain scans were acquired purposely with a low flip angle resulting in low SNR images for the fully sampled scans. Although we are aware that better image quality for the fully sampled scans can be obtained with a higher excitation flip angle, these SNR-deprived scans provided a good starting point to demonstrate the improvement in image quality resulting from our non-uniform k-space averaging approach.

An application in a clinically relevant sequence is the T2-mapping experiment of the knee we performed, shown in Fig. 9. The images acquired with a high T2 weighting are naturally SNR-deprived. We noticed that for the higher T2-prepared acquisitions, which suffer from low SNR, CS-VDA recovered details that were lost in the noise in the fully sampled acquisition. Figure 9B shows an example of an image detail that is being lost in the fully sampled acquisition, for decreasing SNR. This behavior could lead to the loss of high T2 values in a T2-map. The CS-VDA T2 map showed regions of high T2, and a sharp delineation of muscle, which were not visible on the fully sampled T2-map. Furthermore, we saw an increase in SNR in all acquired images, and the calculated T2-map, when using CS-VDA. While the signal shows a convincing decay curve, and a T2 estimate that is in agreement with both methods, the noise behavior is not as straightforward, owing to the optimized $l_1$ constraints. Therefore, the noise is not necessarily constant over the decay times, which reflects on the SNR values. In this study, as in many others, it proved difficult to find suitable objective measures to compare image quality. The insufficiency of objective metrics for image quality is a well-known problem in the CS literature[28]. Frequently-used measures like the structural-similarity index or the mean squared error did not work well in our study because these are strongly biased by the noise characteristics of the images and there is no gold-standard image for comparison for the in-vivo data. For this reason, we chose to evaluate image sharpness as the main quality measure by determining the width of the air/grapefruit-skin interface.

The effects on the SNR arising from accelerated imaging are well understood for parallel imaging[29], however, the noise penalty in CS is more complicated for several reasons. First, the spatial distribution of sampling points in a typical CS measurement gives rise to colored noise[30]. Secondly, $l_1$ regularization inherently leads to denoising, the effect of which depends greatly on the chosen regularization parameter and which makes it difficult to quantify SNR in CS reconstructed images. While application of CS in MRI is one of the most promising applications of this technique, research into the effects of noise on CS reconstruction has been mostly limited to the mathematical literature [31]–[33].

The full potential and flexibility in compressed sensing sampling is not utilized. The link between the sparsifying transform and the sampling transform can be used to design sampling methods that include averaging of k-space points. These insights can be valuable in other imaging fields where CS is used with binary sampling, such as fluorescence microscopy [34], and other Fourier-based sampling modalities, such as radio interferometry [35].

In conclusion, we implemented 3D Cartesian k-space undersampling with a variable number of k-space averages. Additionally, we incorporated the weighted $l_2$-norm into a CS reconstruction. We have shown that variable center-dense k-space averaging outperforms fully sampled k-space sampling for low-SNR MRI acquisitions. We think this novel approach will facilitate improved image quality of inherently low SNR data, such as those with high-resolution or specific contrast-weightings with low SNR efficiency.